\documentclass[runningheads]{llncs}

\usepackage{graphicx}
\usepackage{url}
\usepackage{multirow}
\usepackage{tabularx}

\usepackage{listings}

\newcommand{\name}{DBkWik\texttt{+}\texttt{+}}
\begin{document}

\title{\name - Multi Source Matching of Knowledge Graphs}

\author{Sven Hertling\orcidID{0000-0003-0333-5888} \and
Heiko Paulheim\orcidID{0000-0003-4386-8195}}

\institute{Data and Web Science Group, University of Mannheim, Germany\\
\email{\{sven,heiko\}@informatik.uni-mannheim.de}}

\maketitle

\begin{abstract}

Large knowledge graphs like DBpedia and YAGO are always based on the same source, i.e., Wikipedia.
But there are more wikis that contain information about long-tail entities such as wiki hosting platforms like Fandom.
In this paper, we present the approach and analysis of \name, a fused Knowledge Graph from thousands of wikis.
A modified version of the DBpedia framework is applied to each wiki which results in many isolated Knowledge Graphs.
With an incremental merge based approach, we reuse one-to-one matching systems to solve the multi source KG matching task.
Based on this alignment we create a consolidated knowledge graph with more than 15 million instances.

\keywords{Knowledge Graph Matching \and Incremental Merge \and Fusion}
\end{abstract}

\section{Introduction}

There are many knowledge graphs (KGs) available in the Linked Open Data Cloud\footnote{\url{https://lod-cloud.net}}, some have a special focus like life science or governmental data. In the course of time, a few hubs evolved which have a high link degree to other datasets -- two of them are DBpedia and YAGO. Both cover general knowledge which is extracted from Wikipedia. Thus many applications use these datasets. One drawback is that they all originate from the same source and have thus nearly the same concepts. For many applications like recommender systems~\cite{recommender}, information about not so well known entities (also called \emph{long tail entities}) is required to find similar concepts. Additional sources for such entities can be found in other wikis than Wikipedia.

One example is the wiki farm \emph{Fandom}\footnote{\url{https://www.fandom.com}}. Everyone can create wikis about any topic. Due to the restricted scope in each of these wikis, also pages about not so well known entities are created. As an example William Riker (the fictional character in the Star Trek universe) appears in Wikipedia because this character is famous enough to be added (see also the notability criterion for people\footnote{\url{https://en.wikipedia.org/wiki/Wikipedia:Notability_(people)}}). For other characters, like his mother Betty Riker, this notability is not given, so it only appears in special wikis like \emph{memory alpha}\footnote{\url{https://memory-alpha.fandom.com/wiki/Betty_Riker}} (a Star Trek wiki in Fandom).

The idea in this work is to use these wikis and apply a modified version of the DBpedia extraction framework to create knowledge graphs out of them containing information about long tail entities. Each resulting KG is isolated and can share same instances, properties, and classes that need to be matched.
For this multi source knowledge graph matching task we reuse a one-to-one matcher and apply it multiple times in an incremental merge based setup to create an alignment over all KGs.

After fusing all KGs together based on the generated alignment, we end up with \name, a fused knowledge graph of more than 40,000 wikis from Fandom.
The contributions of this paper are threefold:
\begin{itemize}
\item presentation of the overall approach to generate the KGs
\item matching of 40,000 KGs on schema and instance level by reusing a one-to-one matcher
\item analysis of the resulting knowledge graph
\end{itemize}

The rest of this paper is structured as follows. Section~\ref{sec:relatedwork} describes related work such as different general purpose knowledge graphs and matching techniques.
Afterward, section~\ref{sec:dbkwik} shows details about the retrieval of wikis, application of the DBpedia extraction framework, and the incremental merge of the KGs. After the fusion and provenance information of the KG, the resulting alignment is profiled in section \ref{sec:analysis_alignment} and the KG in section \ref{sec:analysis_kg}. We conclude with an outlook and future work.

\section{Related Work}
\label{sec:relatedwork}

\begin{table*}[t]
    \caption{Comparison of public Knowledge Graphs based on~\cite{KGoverview} sorted by the number of instances. }
    \label{tab:KGcompare}
	\centering
	\begin{tabular}{|l|r|r|r|r|r|}
		\hline
        Knowledge Graph & \# Instances & \# Assertions & \# Classes & \# Relations & Source\\
        \hline
        Voldemort       & 55,861      &  693,428     & 621       & 294     & Web \\
        Cyc             & 122,441     &  2,229,266   & 116,821   & 148     & Experts \\
        DBpedia         & 5,044,223   &  854,294,312 & 760       & 1,355   & Wikipedia \\
        NELL            & 5,120,688   &  60,594,443  & 1,187     & 440     & Web  \\
        YAGO            & 6,349,359   &  479,392,870 & 819,292   & 77      & Wikipedia \\
        CaLiGraph       & 7,315,918   &  517,099,124 & 755,963   & 271     & Wikipedia \\
        BabelNet        & 7,735,436   &  178,982,397 & 6,044,564 & 22      & multiple \\
        DBkWik          & 11,163,719  &  91,526,001  & 12,029    & 128,566 & Fandom \\
        \textbf{\name}  & 15,346,033  &  106,347,347 & 15,642    & 215,273 & Fandom \\
        Wikidata        & 52,252,549  &  732,420,508 & 2,356,259 & 6,236   & Community \\
        \hline
	\end{tabular}
\end{table*}

This section is divided into two parts: 1) description of other cross-domain knowledge graphs and 2) multi source matching approaches to combine isolated KGs into one large KG.

Table~\ref{tab:KGcompare} shows the public cross-domain KGs together with the number of instances, assertions, classes, and relations. In addition, the main source of the content is provided in the last column. The KGs are sorted by the number of instances starting with the smallest.

VoldemortKG~\cite{voldemort} uses data extracted from webpages (common crawl) via structured annotations using
approaches like RDFa, Microdata, and Microformats. The resulting set of KGs (one for each webpage) is then merged together and linked to DBpedia by using Wikipedia links occurring on those webpages. The overall graph is relatively small and contains only 55,861 instances.

Cyc~\cite{cyc} was generated by a small number of experts. They focus on common sense knowledge and create more assertions than instances.
The scalability is quite limited because of the manual generation. The total cost was estimated as 120 Million USD. The numbers in the table refer to the openly available subset OpenCyc.

DBpedia~\cite{dbpedia} instead uses another approach that scales much better. The main source is Wikipedia where a lot of entries contain information in so-called infoboxes (MediaWiki templates). These templates contain attribute value pairs where the values are shown on the webpage.
When processing these pages without resolving the templates, those key-value pairs can be extracted and transformed to triples where the wiki page is the subject, the template key is the property, and the template value is the corresponding literal or resource (in case it is a URL). This scales much better and opens the door for other data-driven approaches.

NELL~\cite{nell} (Never-Ending Language Learning) is an approach to extract information from free text appearing on web pages.
Based on some initial facts, textual patterns for these relations are extracted and applied to unseen text to extract more subjects and objects for that relation. The resulting facts are again used to derive new patterns. With a human-in-the-loop approach, the authors try to increase the quality by removing incorrect triples.

YAGO~\cite{yago} uses the same source as DBpedia (namely Wikipedia) but creates the class hierarchy based on the categories defined in Wikipedia instead of manually creating the hierarchy like DBpedia.

CaLiGraph~\cite{caligraph} also uses the category tree but converts the information in category names into formal axioms e.g. ``List of people from New York City`` where each instance in this category should have the triple $<$instance, bornIn, New York City$>$. 

Babelnet~\cite{babelnet} integrates a lot of sources like Wikipedia and WordNet~\cite{wordnet} to collect synonyms and translations in many languages.

DBkWik~\cite{dbkwik} is generated from Fandom wikis with the DBpedia extraction framework. Thus it has the same structure as DBpedia but includes more long-tail entities, especially from the entertainment domain. It uses information from 12,840 wikis.

Wikidata~\cite{wikidata} is a community-driven approach like Wikipedia but allows to add factual information in form of triples instead of free text.
Furthermore, it includes and fuses other large-scale datasets such as national libraries’ bibliographies.

Many of the above mentioned approaches need some kind of data integration to build up the final KG.
Thus matching algorithms are compared in the following by dividing into entity vs KG matching and two sources vs multi source matching.

In the case of entity matching, the schema is fixed and usually already aligned. If the schema alignment is missing, it can be manually created due to the reduced size of attributes. In KG matching, on the other hand, the concepts are of different classes (e.g. persons, places, events, etc) and can be described using many properties (e.g. name, hair color, height) which in addition can also connect entities (e.g. location property which links a person to a location).
The schema (including classes and properties) is usually not aligned and cannot be created manually due to its large size.

Entity and KG matching exists in two flavors: one to one and multi source matching. In the former case, two inputs are given whereas in multi source matching a set of data sources needs to be matched. Even though the one to one matching can be seen as a special case of multi source matching, many matchers are only able to align two sources at the same time.

There are many one to one entity matching tools available like rule based systems~\cite{rulebasedEM} and deep neural network based ones (DeepMatcher~\cite{deepmatcher}). In addition, configurable frameworks like  MOMA~\cite{moma} are discussed in \cite{kopcke2010frameworks}.
SILK~\cite{silk} matches data sources formatted in RDF but focuses on entities rather than the schema. It uses human generated rules which can be further improved by supervised methods.
For multi source entity matching there exists matching systems like FAMER~\cite{famer} and ALMSER~\cite{almser} which in addition uses active learning to reduce the size of the initial alignment.

For one to one KG matching, the ontology alignment evaluation initiative (OAEI \cite{oaei2021}) allows submitting ontology and KG matching systems which are evaluated by the track/dataset organizers. Starting in 2018, the OAEI contains also a KG track which requires instance as well as schema matches~\cite{oaeiKG}. All systems participating in this track are able to produce a one to one KG alignment like the top performing systems AML~\cite{aml} and ALOD2Vec~\cite{alod2vec}. 
For multi source KG matching~\cite{ordermatters} reuses one to one matching systems to solve the multi source task.
\cite{saeedi2021matching} focus on matching entities in multiple KGs using hierarchical agglomerative clustering (HAC).

%In the following, we discuss the generation of \name.

\section{\name}
\label{sec:dbkwik}

\begin{figure}[t]
    \centering
    \includegraphics[width=\textwidth]{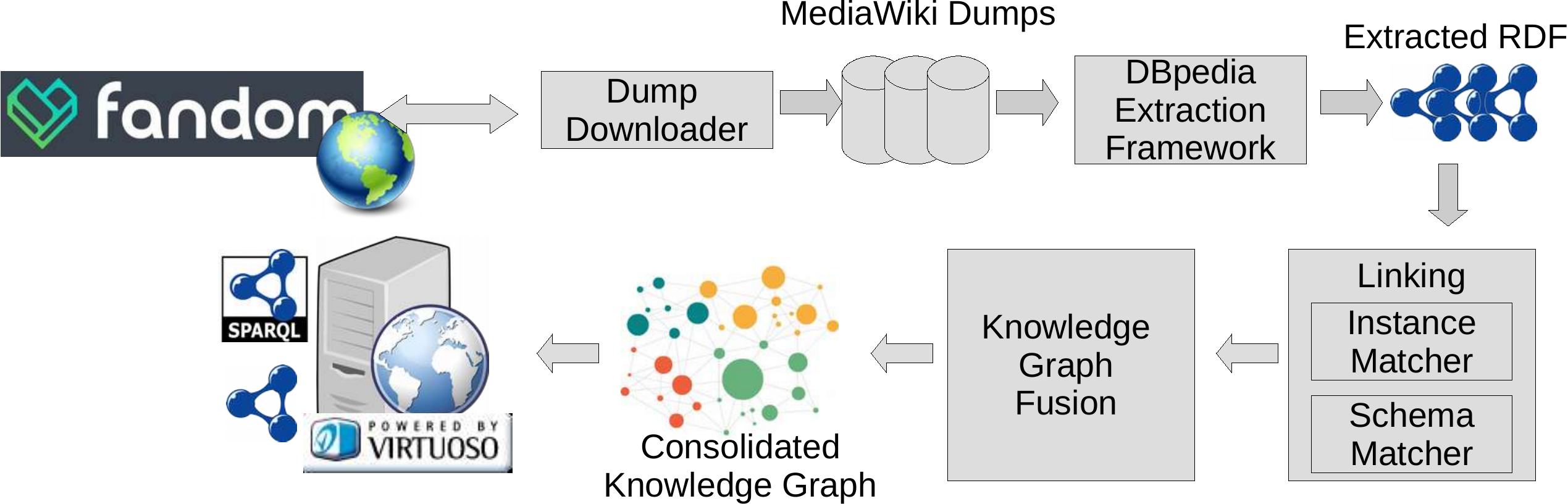}
    \caption{Overview of the approach to create the knowledge graph.}
    \label{fig:overview}
\end{figure}

The overall workflow for creating DBkWik++ is shown in figure \ref{fig:overview}.
First, all wiki dumps need to be downloaded from Fandom. Afterward, a modified version of the DBpedia extraction framework is applied to extract information in RDF. The result is many isolated KGs which are linked and fused to create a consolidated KG. 

\subsection{Acquisition of Wiki Dumps}

The wiki hosting platform Fandom allows wiki creators to provide dumps of their wikis once it is manually triggered\footnote{\url{https://community.fandom.com/wiki/Help:Database_download}}. These files can be downloaded easily. Unfortunately, not many creators do it and thus the coverage of wikis to download is rather low. It can also be very outdated depending on the time the dump is initiated. Therefore, we reached out to Fandom to ask for another way of getting these dumps, but did not succeed. As an alternative solutions, all dumps were downloaded by the wikiteam\footnote{\url{https://github.com/WikiTeam/wikiteam}} software which archives MediaWikis.
This resulted in 307,466 dumps (208,552 of them in English).

\subsection{Creating KGs with DBpedia Extraction Framework}

The wiki dumps generated in the previous steps consist of MediaWiki markup and need to be transformed to a KG. For this step, a modified version of the DBpedia extraction framework is used.
As the name suggests, it is used to create the DBpedia~\cite{dbpedia} knowledge graph.
The transformation of a wiki to a KG works as follows: each wiki page corresponds to a resource and each infobox\footnote{a template in MediaWiki which usually contains the text \texttt{infobox} to visualize important information at the top right corner of a page} becomes a class. Due to the fact that templates like infoboxes do not form a hierarchy, the classes do not have a hierarchy as well. In DBpedia, this hierarchy is created manually in the corresponding mappings wiki\footnote{\url{http://mappings.dbpedia.org}}. 
Similarly, properties are homogenized in case the attribute key of the template is different e.g.
the template properties \texttt{birthdate}, \texttt{birth\_date}, and \texttt{dateofbirth} are mapped to the ontological property \texttt{birthDate}.
All extractors in the DBpedia extraction framework which use knowledge of the mappings wiki are excluded because no mapping exists for Fandom and a manual creation would not scale to hundreds of thousands of wikis.
Instead, the extractors which do not make use of the mapping are included e.g. types and properties are created from infobox template names and infobox keys.
In the usual DBpedia extraction, the abstracts of a wiki are extracted by setting up a MediaWiki instance and calling an API. This is replaced by using another extractor based on the Sweble parser~\cite{sweble} to decrease the runtime.

\subsection{Linking}

\begin{figure}[t]
    \centering
    \includegraphics[width=0.7\textwidth]{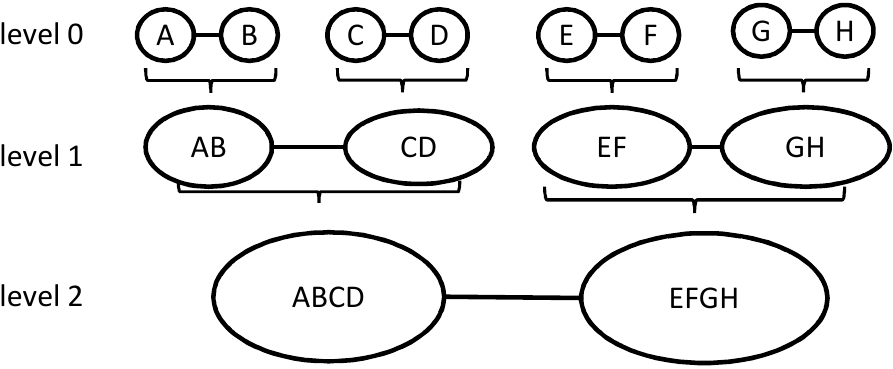}
    \caption{Incremental merge strategy for wikis A to H.}
    \label{fig:incremental}
\end{figure}

The output of the previous step is an isolated knowledge graph per wiki. It happens a lot that the same concept appears in multiple wikis, e.g., the actor Tom Cruise appears in moviepedia wiki, topgun wiki, and jack reacher wiki.
To create a consolidated KG, all source KGs need to be matched together. This does not only require instance matches but also schema level correspondences which include property and class matches. Thus, the KG matching task at hand is different from entity matching because the schema in each data source can be quite different.
The reason is that each attribute value pair in the infobox is transformed into a triple. Therefore, different information about a resource might be extracted e.g. eye color of Tom Cruise in jack reacher wiki and height in topgun wiki. Matching systems needs to be very precise and should be able to handle multi source KG matching.
A blocking approach based on entity types is not possible because of the missing schema alignment.

Due to the fact that there are not many systems able to solve the multi source KG matching task, we implemented an approach presented in~\cite{ordermatters}. The basic idea is to use a one to one matching system over and over again in a multi source setup.
The implementation of the all-pairs approach does not scale (which runs the binary matching system for all KG combinations). For $n$ KGs this would need $\frac{n * (n-1)}{2}$ executions of the matching system e.g. $\frac{300,000 * 299,999}{2}=44,999,850,000$ comparisons. Even if each comparison of two KGs only needed one second (which is overly optimistic), the whole computation would require 1,426 years without parallelization. The same complexity also applies to the resulting mapping size. If each correspondence is stored in a file, it becomes very huge even for a small number of wikis.

Thus the selected approach is called incremental merge based on similarity. Figure \ref{fig:incremental} shows an example for isolated KGs \emph{A} to \emph{H}. The one to one matcher is first applied to KG \emph{A} and \emph{B} which results in an alignment. This alignment is used to create a union KG \emph{AB} out of it. It is generated by adding all triples of \emph{A} and \emph{B} to the result but replacing URIs in \emph{B} with the corresponding one from \emph{A} in case it exists. Thus more information about an entity is available at higher levels of the execution.

The order in which those KGs are merged is determined by the similarity between them. The overall tree is computed by an agglomerative hierarchical clustering (HAC) using tf-idf vectors as features to lower the impact of frequent terms, e.g. upper-level ontology terms. The vectors are generated by using all values which occur in literals and are string-valued. Creating tokens is done in a standard way by sentence splitting, tokenization,
lowercasing, stopword removal, and stemming. The tokenization is adapted due to special naming conventions (like underscore between words or camel case). The output of the hierarchical clustering can be directly used as an execution plan for the matching and merging of KGs.
The advantage of such an approach is that the schema and instances are matched at the same time, at least if the binary matching system is able to do so.

The hierarchical agglomerative clustering algorithm works in the following way: (1) compute the distance matrix (2) let each datapoint be a cluster (3) repeatedly merge the two closest clusters and update the distance matrix accordingly until only a single cluster remains. 
It turned out that many implementations like ELKI~\cite{elki} and SMILE~\cite{Smile} do not scale to more than 65,535 examples (and in this case data sources) because the distance matrix is stored in one array which can only hold up to 2,147,483,647 values (whereas the lower triangular matrix needs $\frac{n * (n-1)}{2}$ entries).
Thus we use the largest 40,000 wikis for merging. The main time is spent in part (1) where the distance matrix is computed (799,980,000 values for 40,000 input wikis). Therefore a faster way is implemented which runs in parallel and uses low-level BLAS calls to compute multiple distances at the same time.

One important parameter is the linkage criteria which heavily influences the height of the resulting tree.
All executions within one level (shown in figure \ref{fig:incremental}) can be executed in parallel because they do not depend on each other. Thus the height corresponds to the number of matching stages which need at least to be executed sequentially. When using single linkage, the corresponding height is 22,007 with 5,869 possible parallel merges in level 0. This is in contrast to complete linkage where the height is only 145 and 12,639 merges in level 0. Thus complete linkage is chosen because of the improved parallelism. In addition, an efficient algorithm for computing complete linkage in $O(n^2)$ exists which is called CLINK~\cite{clink}.

Choosing a one to one matching system is done by analyzing the results of OAEI 2021~\cite{oaei2021}.
Alod2Vec\cite{alod2vec} was one of the top performing systems in the KG track which is very similar to the KGs used in this work. Furthermore, the system is able to produce schema as well as instance alignments.
A postprocessing step is added to the matching system to ensure a $1:1$ alignment which is required by the multi source matching strategy in each execution step. For this, a naive descending approach~\cite{mappingextraction} is used which sorts the alignment by confidence and extracts the correspondences as long as no duplicate source or target entities appear.
The overall runtime of the matching needs six days - most probably due to costly matching approaches executed by the one to one matcher. In the output alignment, no property mappings are generated. As a replacement, the property alignment of a simpler element-based matcher is used.

Even though a blocking-based approach for matching would be possible, each entity combination in the identity set needs to be analyzed and thus the information about the schema and instances of each KG needs to present. 

\subsection{Knowledge Graph Fusion}

Given the alignment and the source KGs, this section describes the approach to generate a fused KG.
The basic idea is to replace the URIs of matched entities with a canonical URI representing the set of identical resources.

Due to the fact that the alignment is generated on various levels and in each of them, only correspondences between two concepts are made, the transitive closure of the alignment needs to be computed. As an example, a concept $a_{1}$ from KG $A$ is mapped to concept $b_{1}$. In the resulting union $AB$, all appearances of $b_{1}$ are replaced with $a_{1}$. In the following, the union is matched to $CD$ which results in the correspondence $<a_{1}, c_{1}>$. Only under the transitive closure $b_{1}$ and $c_{1}$ are also matched. For computing the transitive closure the approach of \cite{transitiveClosure} (Section 3.3) is re-implemented to store the closure in memory which decreases the runtime.

Given the identity sets computed by the transitive closure, 
the next step is to calculate for each set the corresponding canonical URI which is the replacement for all URIs in this set. In contrast to Wikidata, the URIs should have speaking names. The transitive closure is thus sorted by descending size (number of elements in this set). Starting from the largest set, the following algorithm is applied: the most common URI fragment is used as the canonical URI. In case the most common fragment is already in use, a postfix with an increasing number is added. This results in URI fragments like New\_York, New\_York\_1, New\_York\_2 to distinguish between different cities (but just in the case that the majority of the fragments are New\_York).
For new extractions of the knowledge graph, one should make sure that the URIs are stable and always pointing to the same disambiguated entity. Thus it should be checked if the URIs of the new extraction correspond to the same entity as in the older extraction (which usually requires an additional matching step).

In addition to merging the KG, different sources may provide conflicting data values for specific properties~\cite{buildKG}. 
In data integration frameworks like WInte.r~\cite{winter}, conflict resolution functions are provided on an attribute basis e.g. for property first\_name use the shortest (or longest) string. Due to the number of relations in a KG, such an approach does not scale. Currently, a union-based approach is used for all properties.

\subsection{Provenance Information}
\label{sec:prov}

\begin{figure}[t]
    \centering
    \includegraphics[width=0.8\textwidth]{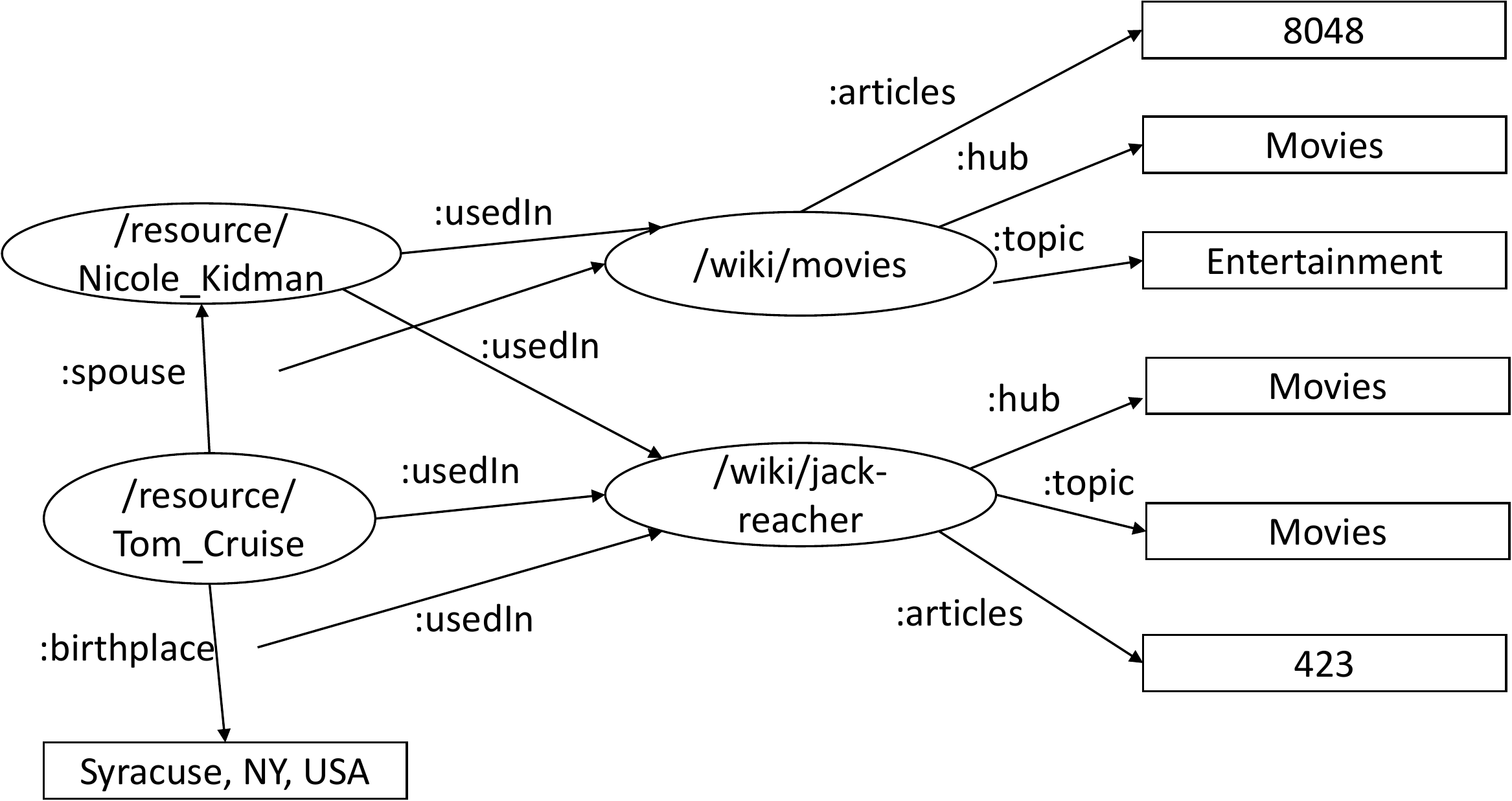}
    \caption{Provenance information such as wiki information about resources.}
    \label{fig:provenance}
\end{figure}

After the fusion of all KGs, the information which resource originates from which wiki is lost.
Thus each wiki is represented in the final KG as its own resource with additional metadata attached to it.
It is collected during the crawling of the wikis. Each MediaWiki has a special statistics page that lists information about the number of pages, users, active users, etc. In Fandom, additional statistics like the WAM score are created. It is a value between 0 and 100 and should indicate the activeness of a wiki\footnote{\url{https://community.fandom.com/wiki/WAM_Hall_of_Fame}}. The computation is based on page views, contributions, and level of user activity. All this information is added to the resource representing the corresponding wiki. Figure~\ref{fig:provenance} shows an example where the property \textit{usedIn} connects the resources Nicole Kidman and Tom Cruise to the corresponding source wikis. Note that one resource (like Nicole Kidman in this example) might be extracted from multiple wikis.

After this step, the final consolidated KG is generated. In the following two sections, we analyze the resulting alignment as well as the final KG.

\section{Analysis of the Resulting Alignment}
\label{sec:analysis_alignment}

The generated alignment consists of 7,847,480 correspondences on the schema and instance level.
Out of those, 5,254,001 correspondences have the same label and 2,593,479 correspondences (33.05\%) link entities that do not share a label. 
The fraction of 33\% is in line with the reference alignment of the OAEI KG track which has between 16\% and 47\% matches of entities with different labels.~\cite{oaeiKG}.

\begin{table}[t]
	\centering
	\caption{Statistics about the identity sets generated by the transitive closure over the alignment}
	\label{tab:stats_align}
	\begin{tabular}{|l|r|r|r|r|r|}
\hline
           & \#Identity Sets         & Min \#Elements & Max \#Elements & Mean  & Std Dev \\ \hline
Classes    & 22,772    & 2              & 11,899         & 12.83 & 139.53  \\ \hline
Properties & 54,814    & 2              & 16,929         & 21.90 & 219.40  \\ \hline
Instances  & 2,097,598 & 2              & 821            & 4.07  & 5.26    \\ \hline
\end{tabular}
\end{table}

\begin{figure}[t]
    \centering
    \includegraphics[width=\textwidth]{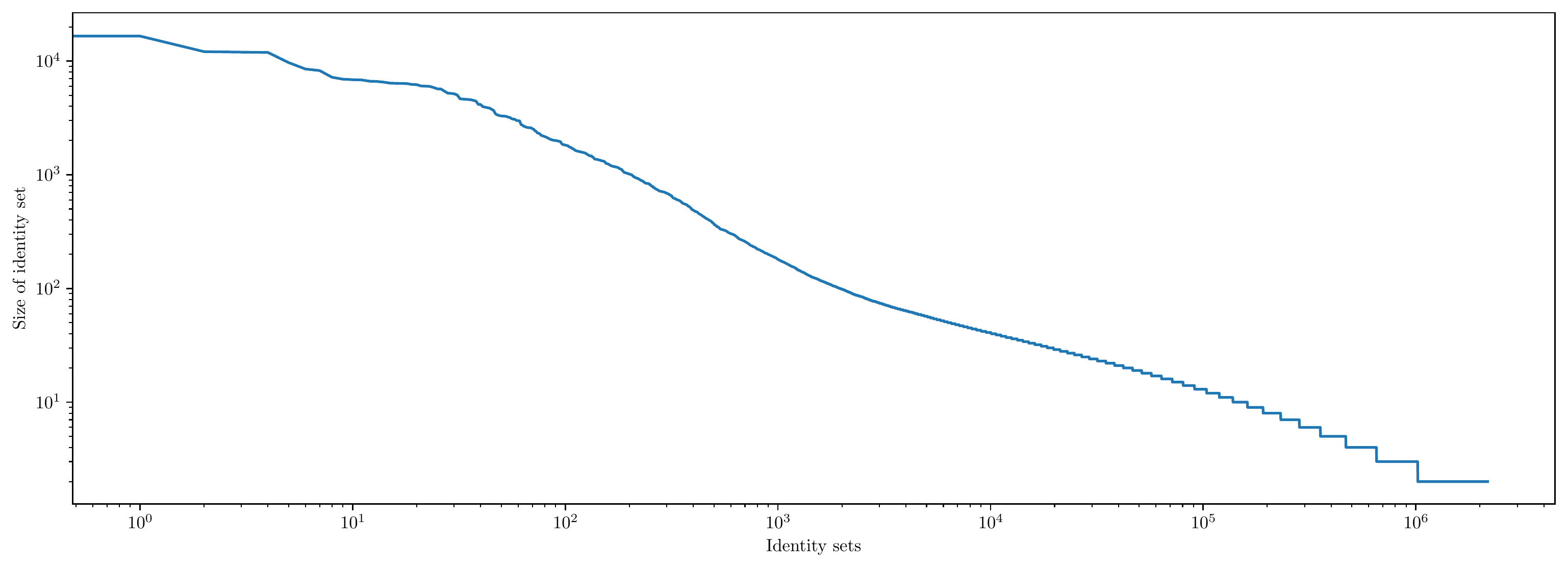}
    \caption{Log-log plot of the number of elements in each identity set sorted by size.}
    \label{fig:closureSize}
\end{figure}

Each correspondence maps only one entity to another and thus the transitive closure needs to be computed as discussed before. It has overall 2,175,184 identity sets. The smallest identity set is of size two and the maximum size is 16,929. Figure \ref{fig:closureSize} shows the number of elements in each identity set in a log-log plot. It shows that there are a few sets with many elements but a lot of sets with a rather low number of identical resources. 

Table \ref{tab:stats_align} shows the statistics divided into classes, properties, and instances.
One can see that properties are matched the most and also contains the largest number of elements in one identity set.
This is followed by classes which also occur in many different KGs.
Due to the restriction that one entity can be matched to at most one other entity in another KG, the number of elements in an identity set also represents the number of KGs in which a concept can be found.
Table \ref{tab:application} shows the evaluation measures for a very small subset of the alignment where a gold standard is given by the OAEI track.

\begin{table}[t]
	\centering
	\caption{Most frequently matched classes, properties, and instances}
	\label{tab:most_frequent}
	%\resizebox{\textwidth}{!}{
	\begin{tabular}{|l|r||l|r||l|r|}
		\hline
		\multicolumn{2}{|c||}{Classes} & \multicolumn{2}{c||}{Properties} &   \multicolumn{2}{c|}{Instances}   \\ \hline
		name      &            \#match & name       &             \#match & name                     & \#match \\ \hline
		character &              11,899 & title      &               16,929 & Discussion               &     821 \\
		episode   &               5,153 & name       &               16,593 & Jim Cummings             &     210 \\
		quote     &               4,148 & gender     &               12,109 & Circe                    &     198 \\
		location  &               3,431 & caption    &               11,963 & Rhea                     &     186 \\
		item      &               2,187 & image      &                9,696 & Tara Strong              &     178 \\
		book      &               1,735 & type       &                8,247 & Wonderland               &     177 \\
		season    &               1,564 & height     &                6,931 & Kevin Michael Richardson &     176 \\
		game      &               1,429 & status     &                6,860 & Fred Tatasciore          &     162 \\
		album     &               1,385 & occupation &                6,827 & Tom Kenny                &     160 \\
		film      &               1,325 & species    &                6,655 & Israel                   &     159 \\ \hline
	\end{tabular}
	%}
\end{table}

\begin{table}[t]
\centering
\caption{Evaluation of the multi source matching strategy using the OAEI KG track. For each category, precision (P), recall (R), and f-measure (F) are reported.}
\label{tab:application}

\begin{tabular}{|l|crr|crr|crr|crr|}
\hline
 &
  \multicolumn{3}{c|}{Instance} &
  \multicolumn{3}{c|}{Class} &
  \multicolumn{3}{c|}{Property} \\ \hline
 &
  \multicolumn{1}{c|}{P} &
  \multicolumn{1}{c|}{R} &
  \multicolumn{1}{c|}{F1} &
  \multicolumn{1}{c|}{P} &
  \multicolumn{1}{c|}{R} &
  \multicolumn{1}{c|}{F1} &
  \multicolumn{1}{c|}{P} &
  \multicolumn{1}{c|}{R} &
  \multicolumn{1}{c|}{F1} \\ \hline
Alod2vec &
  \multicolumn{1}{r|}{.937} &
  \multicolumn{1}{r|}{.390} &
  .551 &
  \multicolumn{1}{r|}{.842} &
  \multicolumn{1}{r|}{.432} &
  .571 &
  \multicolumn{1}{r|}{.702} &
  \multicolumn{1}{r|}{.549} &
  .617 \\ \hline
\end{tabular}%

\end{table}

After this quantitative analysis, the top matched entities (largest identity sets) are also inspected.
Table \ref{tab:most_frequent} shows the top ten matched classes, properties, and instances together with the number of elements in the identity set (in how many wikis this concept is contained).
Wiki-related entities like Main Pages are excluded and the names are the most common label appearing in the corresponding identity set (because entities with different labels can be also matched). 

The class \texttt{character} is contained in 11,899 wikis. This also shows the domain of the majority of the wikis which is entertainment, movies, and games. But there are also more general classes like location, game, and film which appear in at least 1,325 wikis. The largest identity set contains the property \texttt{title} which is used in nearly 17,000 out of the 40,000 inspected wikis. It is directly followed by the property \texttt{name} which is used in 16,593 wikis. These properties are used to give a label to an instance. Gender, height, and occupation are more interesting because it describes some attributes of persons. This is also in line with the top matched instances because they are mostly from the class \texttt{person}.
One can already see that the number of matches is rather low for instances which means that fewer wikis have resources in common but rather describe different topics. The instance Jim Cummings appears in 210 wikis (e.g. disney-mickey-mouse, tarzan, sonic, and starwars) and is an American voice actor.

\section{Analysis of the Resulting Knowledge Graph}
\label{sec:analysis_kg}

\begin{figure}[t]
    \centering
    \includegraphics[width=0.85\textwidth]{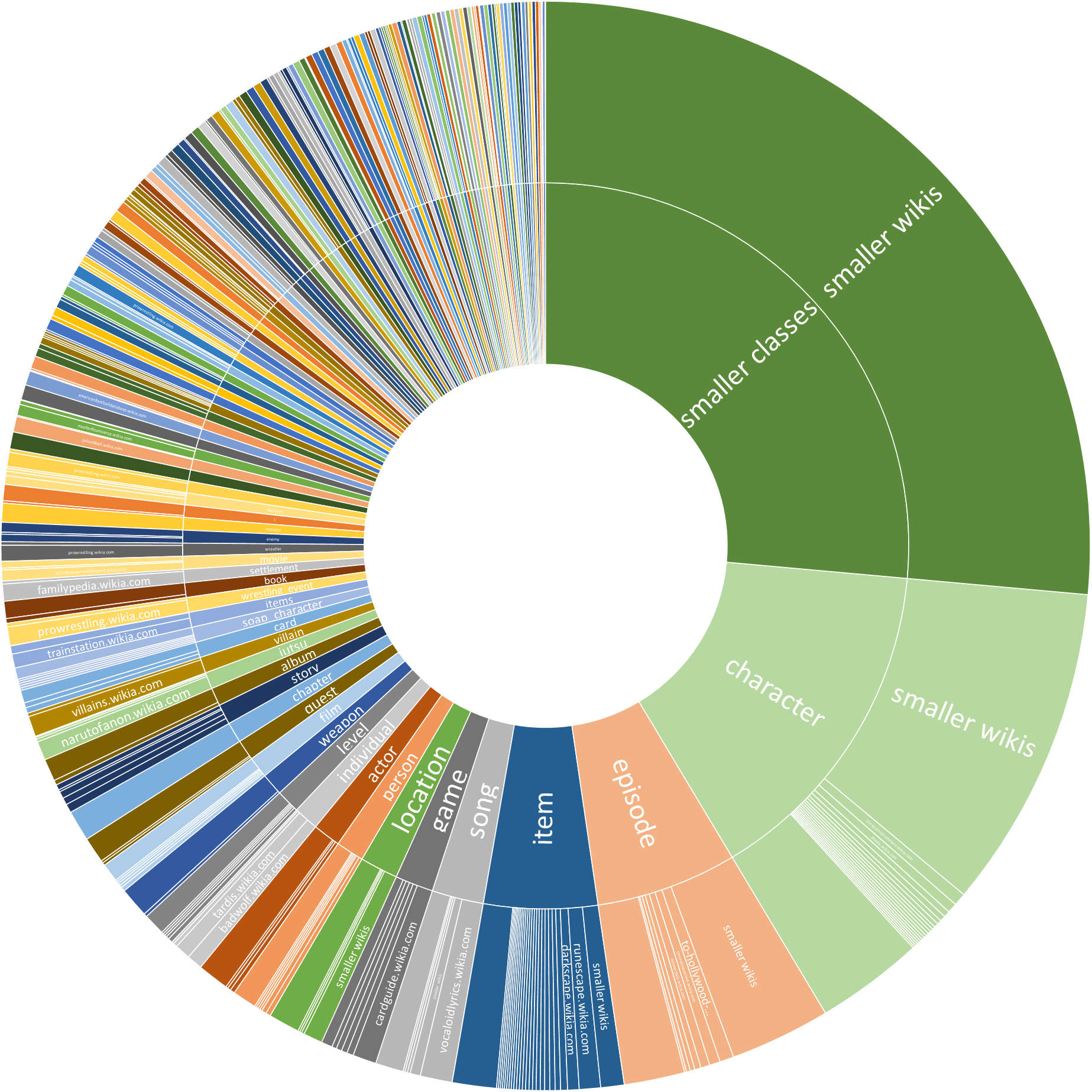}
    \caption{Sunburst diagram of the final KG showing the relative amounts of instances per class. Inner circles: matched classes. Outer circles: Wikis containing this class.}
    \label{fig:sunburst}
\end{figure}

After analyzing the alignment, the resulting fused KG is inspected.
Overall 15,346,033 instances are contained and described by 215,273 properties.
The number of classes is 109,042 but only 15,642 classes are based on infobox templates (which usually better indicate actual classes). The number of assertions depends on which triples are actually counted.
When only using the extracted triples from the WikiMedia templates (main information source), then the value is 106,347,347. When adding additional information like page links, labels, comments, categories, etc the value  increases to 1,032,941,819 triples. Further adding NIF files which provide the structure of wiki pages results in 4,422,408,859 triples. 
Table \ref{tab:KGcompare} shows all reported numbers in comparison to other available KGs like DBpedia, YAGO, and Babelnet. The only slight increase of instances between DBkWik and \name~can be explained by the fact that DBkWik uses the largest 12,000 wikis whereas \name~uses the largest 40,000 wikis and the fact that the size of the wikis follows a power law (meaning there are only a few large wikis but a lot of small wikis).

Figure \ref{fig:sunburst} shows the relative number of instances per matched class via a sunburst diagram.
The inner circle represents the matched classes and the radius corresponds to the number of instances in this class. The outer circle shows the distribution of this class over all wikis containing it.
Due to displaying issues, too small classes/wikis are grouped together (indicated by smaller classes/wikis).
Characters, episodes, and items are the largest classes but on the other side, there are many specific classes e.g. soap character with only a few instances.

\subsection{Resource Availability}
\label{sec:resource}

The final dataset~\cite{resourcedbkwik} can be downloaded from \\ \url{https://doi.org/10.6084/m9.figshare.20407864.v1}.

\section{Conclusion and Outlook}
\label{sec:conclusion}

In this paper, we presented \name, which is a consolidated knowledge graph generated from 40,000 wikis belonging to the Fandom wiki hosting platform. The matching approach reuses top performing binary matchers from the OAEI competition and shows that a multi source matching problem can be reduced to one to one matching tasks in a reasonable time. We analyzed the resulting alignment as well as the final KG to show which kind of information can be found in \name. Due to the fact that it does not originate from Wikipedia, it contains a lot of instances which do not appear there (long tail entities). 

As an outlook, three further steps to improve and evaluate \name~are described in the following paragraphs.

In the current approach, the classes are created from infobox templates. As already discussed, those do not form a hierarchy.
But there are also categories which are manually created, form a hierarchy, and are used to group wiki articles.
Thus it is also possible to use them as a class replacement. \cite{tifi} already clean those category trees to induce a class hierarchy. After this step, each KG has a class hierarchy which needs to be merged to create a huge consolidated hierarchy which does not need to be manually created (like in DBpedia).

For the fusion of the KGs, a union-based approach is currently applied. But for functional properties (only one value per instance) multiple sources can provide conflicting values. A first step toward this direction is to determine if properties are functional. One trivial approach is data driven and checks in each KG if the corresponding property (which is matched across multiple KGs) is functional. If this is the case, then a datatype specific resolution method can be applied (e.g. average for numeric values and shortest/longest values for strings). In the case of a voting-based approach, metadata of the sources (see section \ref{sec:prov}) can be used to give higher weights for the voting e.g. number of editors (more editors in a wiki should increase the trustworthiness of values). A property based resolution method is not suitable because this would require knowledge on how to resolve the values for each property. This would not scale to the presented use case.

A next step is the accuracy evaluation of the resulting knowledge graph with intrinsic and extrinsic methods.
For an intrinsic evaluation, random samples of triples from the knowledge graph can be assessed.
The more of these triples are correct, the more accurate and better is the graph.
This can be scaled to many judges in a crowdsourced survey. An important part is to show the source wiki page from which the information is extracted because the information might be very domain specific and not known to the survey participant.
In an extrinsic evaluation, the knowledge graph is used as an external source in other applications such as
recommender systems or classification approaches.
\cite{noia2016sprank} shows that the coverage of entities in DBpedia for a recommender system is not larger than 85 \% for movies, 63 \% for music artists, and 31 \% for books. The LyricWiki\footnote{\url{https://lyrics.fandom.com}}, which is contained in \name, could help in this scenario to increase the recall for artists. 

\bibliographystyle{splncs04}
\bibliography{bibliography}

\begin{thebibliography}{10}
\providecommand{\url}[1]{\texttt{#1}}
\providecommand{\urlprefix}{URL }
\providecommand{\doi}[1]{https://doi.org/#1}

\bibitem{recommender}
Alshammari, G., Jorro-Aragoneses, J.L., Kapetanakis, S., Petridis, M.,
  Recio-Garc{\'i}a, J.A., D{\'i}az-Agudo, B.: A hybrid cbr approach for the
  long tail problem in recommender systems. In: Case-Based Reasoning Research
  and Development (2017)

\bibitem{dbpedia}
Auer, S., Bizer, C., Kobilarov, G., Lehmann, J., Cyganiak, R., Ives, Z.:
  Dbpedia: A nucleus for a web of open data. In: The semantic web, pp.
  722--735. Springer (2007)

\bibitem{transitiveClosure}
Beek, W., Raad, J., Wielemaker, J., van Harmelen, F.: sameas.cc: The closure of
  500m owl: sameas statements. In: {ESWC} (2018)

\bibitem{tifi}
Chu, C.X., Razniewski, S., Weikum, G.: Tifi: Taxonomy induction for fictional
  domains. In: The World Wide Web Conference (2019)

\bibitem{clink}
Defays, D.: An efficient algorithm for a complete link method. The Computer
  Journal  (1977)

\bibitem{sweble}
Dohrn, H., Riehle, D.: Design and implementation of the sweble wikitext parser:
  unlocking the structured data of wikipedia. In: Proceedings of the 7th
  International Symposium on Wikis and Open Collaboration (2011)

\bibitem{yago}
Fabian, M., Gjergji, K., Gerhard, W., et~al.: Yago: A core of semantic
  knowledge unifying wordnet and wikipedia. In: 16th International world wide
  web conference, WWW (2007)

\bibitem{aml}
Faria, D., Lima, B., Silva, M.C., Couto, F.M., Pesquita, C.: {AML} and {AMLC}
  results for {OAEI} 2021. In: Ontology Matching Workshop at ISWC (2021)

\bibitem{buildKG}
Fensel, D., {\c{S}}im{\c{s}}ek, U., Angele, K., Huaman, E., K{\"a}rle, E.,
  Panasiuk, O., Toma, I., Umbrich, J., Wahler, A.: How to Build a Knowledge
  Graph, pp. 11--68. Springer (2020)

\bibitem{KGoverview}
Heist, N., Hertling, S., Ringler, D., Paulheim, H.: Knowledge graphs on the web
  - an overview. In: Knowledge Graphs for eXplainable Artificial Intelligence.
  {IOS} Press (2020)

\bibitem{caligraph}
Heist, N., Paulheim, H.: Uncovering the semantics of wikipedia categories. In:
  International semantic web conference (2019)

\bibitem{dbkwik}
Hertling, S., Paulheim, H.: Dbkwik: extracting and integrating knowledge from
  thousands of wikis. Knowledge and Information Systems  (2020)

\bibitem{oaeiKG}
Hertling, S., Paulheim, H.: The knowledge graph track at {OAEI} - gold
  standards, baselines, and the golden hammer bias. In: The Semantic Web - 17th
  International Conference, {ESWC}2020 (2020)

\bibitem{ordermatters}
Hertling, S., Paulheim, H.: Order matters: Matching multiple knowledge graphs.
  In: {K-CAP} '21: Knowledge Capture Conference, Virtual Event, USA, December
  2-3, 2021 (2021)

\bibitem{resourcedbkwik}
Hertling, S., Paulheim, H.: {DBkWik Plus Plus} (2022).
  \doi{10.6084/m9.figshare.20407864.v1},
  \url{https://figshare.com/articles/dataset/DBkWik_Plus_Plus/20407864}

\bibitem{kopcke2010frameworks}
K{\"o}pcke, H., Rahm, E.: Frameworks for entity matching: A comparison. Data \&
  Knowledge Engineering  \textbf{69} (2010)

\bibitem{winter}
Lehmberg, O., Bizer, C., Brinkmann, A.: Winte.r - {A} web data integration
  framework. In: {ISWC} 2017 Posters {\&} Demonstrations (2017)

\bibitem{cyc}
Lenat, D.B.: Cyc: A large-scale investment in knowledge infrastructure. Commun.
  ACM  (1995)

\bibitem{Smile}
Li, H.: Smile. \url{https://haifengl.github.io} (2014)

\bibitem{mappingextraction}
Meilicke, C., Stuckenschmidt, H.: Analyzing mapping extraction approaches. In:
  OM (2007)

\bibitem{wordnet}
Miller, G.A.: Wordnet: a lexical database for english. Communications of the
  ACM  \textbf{38}(11),  39--41 (1995)

\bibitem{nell}
Mitchell, T., Cohen, W., Hruschka, E., Talukdar, P., Betteridge, J., Carlson,
  A., Dalvi, B., Gardner, M., Kisiel, B., Krishnamurthy, J., Lao, N., Mazaitis,
  K., Mohamed, T., Nakashole, N., Platanios, E., Ritter, A., Samadi, M.,
  Settles, B., Wang, R., Wijaya, D., Gupta, A., Chen, X., Saparov, A., Greaves,
  M., Welling, J.: Never-ending learning. In: AAAI (2015)

\bibitem{deepmatcher}
Mudgal, S., Li, H., Rekatsinas, T., Doan, A., Park, Y., Krishnan, G., Deep, R.,
  Arcaute, E., Raghavendra, V.: Deep learning for entity matching: {A} design
  space exploration. In: {SIGMOD} Conference 2018 (2018)

\bibitem{babelnet}
Navigli, R., Ponzetto, S.P.: Babelnet: The automatic construction, evaluation
  and application of a wide-coverage multilingual semantic network. Artificial
  intelligence  \textbf{193},  217--250 (2012)

\bibitem{noia2016sprank}
Noia, T.D., Ostuni, V.C., Tomeo, P., Sciascio, E.D.: Sprank: Semantic
  path-based ranking for top-n recommendations using linked open data. ACM
  Transactions on Intelligent Systems and Technology (TIST)  \textbf{8}(1),
  1--34 (2016)

\bibitem{alod2vec}
Portisch, J., Paulheim, H.: Alod2vec matcher results for oaei 2021. In: CEUR
  Workshop Proceedings (2022)

\bibitem{oaei2021}
Pour, M.A.N., Algergawy, A., Amardeilh, F., Amini, R., Fallatah, O., Faria, D.,
  Fundulaki, I., Harrow, I., Hertling, S., Hitzler, P., Huschka, M., Ibanescu,
  L., Jim{\'{e}}nez{-}Ruiz, E., Karam, N., Laadhar, A., Lambrix, P., Li, H.,
  Li, Y., Michel, F., Nasr, E., Paulheim, H., Pesquita, C., Portisch, J.,
  Roussey, C., Saveta, T., Shvaiko, P., Splendiani, A., Trojahn, C.,
  Vatascinov{\'{a}}, J., Yaman, B., Zamazal, O., Zhou, L.: Results of the
  ontology alignment evaluation initiative 2021. In: Ontology Matching Workshop
  at ISWC (2021)

\bibitem{almser}
Primpeli, A., Bizer, C.: Graph-boosted active learning for multi-source entity
  resolution. In: International Semantic Web Conference (2021)

\bibitem{saeedi2021matching}
Saeedi, A., David, L., Rahm, E.: Matching entities from multiple sources with
  hierarchical agglomerative clustering. In: KEOD (2021)

\bibitem{famer}
Saeedi, A., Peukert, E., Rahm, E.: Using link features for entity clustering in
  knowledge graphs. In: European Semantic Web Conference (2018)

\bibitem{elki}
Schubert, E., Koos, A., Emrich, T., Z{\"{u}}fle, A., Schmid, K.A., Zimek, A.: A
  framework for clustering uncertain data. Proc. {VLDB} Endow.  (2015)

\bibitem{rulebasedEM}
Singh, R., Meduri, V.V., Elmagarmid, A.K., Madden, S., Papotti, P.,
  Quian{\'{e}}{-}Ruiz, J., Solar{-}Lezama, A., Tang, N.: Generating concise
  entity matching rules. In: {SIGMOD} Conference 2017 (2017)

\bibitem{moma}
Thor, A., Rahm, E.: {MOMA} - {A} mapping-based object matching system. In:
  Third Biennial Conference on Innovative Data Systems Research, {CIDR} 2007,
  Asilomar, CA, USA, January 7-10, 2007, Online Proceedings (2007)

\bibitem{voldemort}
Tonon, A., Felder, V., Difallah, D.E., Cudr{\'e}-Mauroux, P.: Voldemortkg:
  Mapping schema.org and web entities to linked open data. In: The Semantic Web
  -- ISWC 2016 (2016)

\bibitem{silk}
Volz, J., Bizer, C., Gaedke, M., Kobilarov, G.: Silk - {A} link discovery
  framework for the web of data. In: {WWW2009} Workshop on Linked Data on the
  Web (2009)

\bibitem{wikidata}
Vrande{\v{c}}i{\'c}, D., Kr{\"o}tzsch, M.: Wikidata: a free collaborative
  knowledgebase. Communications of the ACM  \textbf{57} (2014)

\end{thebibliography}

\end{document}